\documentclass[12pt, a4paper]{article}
\usepackage{cite}
\usepackage{amsmath,amssymb}
\input{colordvi.tex}
\usepackage{comment}
\usepackage{bm}
\usepackage{url}
\newcommand{\Slash}[1]{{\ooalign{\hfil/\hfil\crcr$#1$}}}

\usepackage{ifpdf}
\ifpdf
  \usepackage{graphicx, hyperref, xcolor}     
\else     
  \usepackage[dvipdfmx]{graphicx, hyperref, xcolor}     
 \fi

\definecolor{rossoferrari}{HTML}{D9073D}
\definecolor{mediumblue}{HTML}{0000CD}
\hypersetup{
setpagesize=false,
bookmarksnumbered=true,%
bookmarksopen=true,%
colorlinks=true,%
linkcolor=rossoferrari,
urlcolor=mediumblue,
citecolor=mediumblue,
}

\begin{document}

\begin{titlepage}

\begin{center}

\hfill UT-20-02\\

\vskip .75in

{\Large \bf
Detecting Light Boson Dark Matter through \\[.3em] Conversion into Magnon
}

\vskip .75in

{\large
So Chigusa$^{(a)}$, Takeo Moroi$^{(a,b)}$ and Kazunori Nakayama$^{(a,b)}$
}

\vskip 0.25in

$^{(a)}${\em Department of Physics, Faculty of Science,\\
The University of Tokyo,  Bunkyo-ku, Tokyo 113-0033, Japan}\\[.3em]
$^{(b)}${\em Kavli IPMU (WPI), The University of Tokyo,  Kashiwa, Chiba 277-8583, Japan}

\end{center}
\vskip .5in

\begin{abstract}

Light boson dark matter such as axion or hidden photon can be resonantly converted into a magnon in a magnetic insulator under the magnetic field, which can be detected experimentally. We provide a quantum mechanical formulation for the magnon event rate and show that the result is consistent with that obtained by a classical calculation. Besides, it is pointed out that the experimental setup of the QUAX proposal for the axion detection also works as a detector of hidden photon dark matter. It has good sensitivity in the mass range around 1\,meV, which is beyond astrophysical constraints.

\end{abstract}

\end{titlepage}


\renewcommand{\thepage}{\arabic{page}}
\setcounter{page}{1}
\renewcommand{\thefootnote}{\#\arabic{footnote}}
\setcounter{footnote}{0}

\newpage

\tableofcontents

\section{Introduction}
\label{sec:Intro}

Light bosonic dark matter (DM) is one of the well-motivated frameworks of DM model~\cite{Jaeckel:2010ni,Arias:2012az}.
The QCD axion is the best-known example~\cite{Preskill:1982cy,Abbott:1982af,Dine:1982ah,Kim:1986ax,Kawasaki:2013ae}, but more general axion-like particles may also be motivated from string theory~\cite{Svrcek:2006yi,Arvanitaki:2009fg,Cicoli:2012sz}.
The axion-like particle can easily have a correct relic abundance through coherent oscillation. It has an (almost) homogeneous field value during inflation, which eventually becomes a coherently oscillating field when the Hubble parameter decreases to the axion mass. It behaves as a non-relativistic matter thereafter.
There are many experimental ideas dedicated to detecting axion DM.
The cavity haloscope~\cite{Sikivie:1983ip,Bradley:2003kg} experiments including ADMX~\cite{Asztalos:2009yp}, HAYSTAC~\cite{Zhong:2018rsr}, ORGAN~\cite{McAllister:2017lkb}, KLASH~\cite{Alesini:2017ifp}, CULTASK~\cite{Semertzidis:2019gkj}, as well as MADMAX~\cite{Horns:2012jf,Jaeckel:2013eha,TheMADMAXWorkingGroup:2016hpc}, ABRACADABRA~\cite{Kahn:2016aff} and also other ideas~\cite{Marsh:2018dlj,Obata:2018vvr,Nagano:2019rbw,Lawson:2019brd,Zarei:2019sva} use the axion-photon coupling of the form $\mathcal L \propto a F_{\mu\nu}\widetilde F^{\mu\nu}$, while the CASPEr~\cite{Budker:2013hfa} uses the axion-nucleon coupling, and the QUAX~\cite{Barbieri:1985cp,Barbieri:2016vwg,Crescini:2020cvl} is sensitive to the axion-electron coupling.
They already exclude broad parameter regions of axion mass and coupling constant and some of them begin to reach the parameter regions predicted by the QCD axion.

The hidden photon is another well-motivated candidate of DM, which is also expected to show up in the string theory framework~\cite{Cicoli:2011yh}.
There are several scenarios for hidden photon DM production with sub-eV mass scale: production through the axionic coupling~\cite{Agrawal:2018vin,Co:2018lka,Bastero-Gil:2018uel}, scalar coupling~\cite{Dror:2018pdh}, cosmic strings~\cite{Long:2019lwl}, inflationary fluctuation~\cite{Graham:2015rva}, gravitational production~\cite{Ema:2019yrd} and coherent oscillation~\cite{Arias:2012az,AlonsoAlvarez:2019cgw,Nakayama:2019rhg}.
There are many experiments dedicated to the detection of hidden photon~\cite{Horns:2012jf,Wagner:2010mi,Parker:2013fxa,Chaudhuri:2014dla,Hochberg:2016ajh,Hochberg:2016sqx,Bloch:2016sjj,Knapen:2017ekk,Griffin:2018bjn,Hochberg:2017wce,Arvanitaki:2017nhi,Baryakhtar:2018doz}.

In this paper, we explore the possibility to detect axion and hidden photon DM.  In particular, the detection of light boson DM may be
possible using the ferromagnet or ferrimagnet insulator through the
magnon (i.e., electron spin wave) excitation.  Such an idea was
proposed for the axion DM detection in the QUAX proposal~\cite{Barbieri:1985cp,Barbieri:2016vwg}.
We point out that the experimental setup of the QUAX also works as a
hidden photon DM detector.  Through the small kinetic mixing with the
Standard Model (SM) photon, the hidden photon interacts with the SM
particles.  The hidden photon DM can excite the magnon in the magnetic
insulator. It can be viewed as a conversion of a hidden photon into a
magnon in the field theory language. By applying the external magnetic
field, the magnon frequency can be tuned so that it matches the DM
mass and the conversion is kinematically accessible.  We compare the
sensitivities of axion and dark photon searches using magnon and those
using cavity mode of the electromagnetic wave in the QUAX-like setup.

In Sec.~\ref{sec:magnon} we briefly review the property of the
magnon. In particular, its dispersion relation is derived. In
Sec.~\ref{sec:conv} the axion DM conversion rate into the magnon is
calculated and the experimental sensitivity is estimated. First, we
derive the axion-magnon conversion rate by a quantum mechanical
calculation, which has advantageous applicability to the case with only a small number of magnons.
Then we apply the same method for the hidden photon DM in Sec.~\ref{sec:hidden}.  In Sec.~\ref{sec:conc} we
discuss another idea to use the (proposed) axion detector as a hidden photon detector.

\section{Magnon in ferromagnetic materials}
\label{sec:magnon}

In the insulator, the outermost electrons bounded by each atomic cell may contribute to the magnetic properties.
For example, in the case of the Yttrium Iron Garnet (YIG) used for the QUAX, five electrons in the outermost orbit of each Iron atom explain its ferromagneticity.
Let us start with the Heisenberg model
\begin{align}
	H = -g\mu_B \sum_\ell \sum_j \vec B^0 \cdot \vec S_{\ell j}
  - \frac{1}{2} \sum_{\ell,\ell'} \sum_{j,j'} \left( J_{\ell \ell' j j'} \vec S_{\ell j} \cdot \vec S_{\ell' j'}
  + \sum_{\alpha \beta} D_{\ell \ell' j j'}^{\alpha\beta} S_{\ell j}^\alpha S_{\ell' j'}^\beta \right),
	\label{Heisenberg}
\end{align}
where $\vec S_{\ell j}$ is the total electron spin at each cell, $g=2$, $\mu_B = e/(2m_e)$ is the Bohr magneton with $e$ and $m_e$ being the absolute electromagnetic charge and mass of the electron, respectively, and $\vec B^0$ is the external magnetic field.
Here, $\ell, \ell' = 1,\dots, N$ labels magnetic unit cells, while $j, j' = 1,\dots, n$ labels atomic cells inside a magnetic unit cell.
Note that the indices $j, j'$ can also be viewed as labels of the sublattice.
The interaction terms proportional to $J_{\ell \ell' j j'}$ and $D_{\ell \ell' j j'}^{\alpha\beta}$ with $\alpha, \beta$ being the vector indices are called the exchange and dipole interactions, respectively.
Hereafter, we neglect dipole interaction since it is typically much smaller than the exchange interaction.

In some species of ferromagnetic materials including the YIG and many other ferrimagnetic insulators, electrons belonging to different sublattices have different directions to which their spins are oriented.
We introduce a local coordinate system for each atomic cell in which the total electron spin $\vec S'_{\ell j}$ is oriented to the $z$ direction in the ground state.
Besides, we introduce $n$ rotation matrices $R_j^{\alpha \beta}$ with which we can relate local and global coordinate systems as
\begin{align}
  S_{\ell j}^\alpha = \sum_{\beta} R_j^{\alpha \beta} S_{\ell j}^{' \beta}.
\end{align}
Here, we define the global coordinate system such that the magnetic moment of the material is along with the $z$ direction.
The rotation matrices $R_j^{\alpha\beta}$ are determined so that the total energy of the system is minimized when $S_{\ell j}^{' \beta} \propto \delta^{\beta z}$.
Thus, the explicit form of $R_j^{\alpha\beta}$ depends on the details of the material, namely, the relative size and sign of exchange interactions $J_{\ell\ell'jj'}$.

It is convenient to consider fluctuations around the ground state with creation and annihilation operators introduced by the Holstein-Primakoff transformation
\begin{align}
	&S_{\ell j}^{'+} \equiv S_{\ell j}^{'x} + i S_{\ell j}^{'y} = \sqrt{2s_j} \sqrt{1-\frac{\widetilde c_{\ell j}^\dagger \widetilde c_{\ell j}}{2s_j}} \widetilde c_{\ell j},\\
	&S_{\ell j}^{'-} \equiv S_{\ell j}^{'x} - i S_{\ell j}^{'y} = \sqrt{2s_j} \widetilde c_{\ell j}^\dagger \sqrt{1-\frac{\widetilde c_{\ell j}^\dagger \widetilde c_{\ell j}}{2s_j}},\\
	&S_{\ell j}^{'z} = s_j - \widetilde c_{\ell j}^\dagger \widetilde c_{\ell j},
\end{align}
where $s_j$ is the total spin at each cite belonging to the sublattice $j$, which takes a universal value of $5/2$ for the YIG, and
\begin{align}
	\left[ \widetilde c_{\ell j}, \widetilde c_{\ell' j'}^\dagger \right] = \delta_{\ell\ell'} \delta_{jj'}.
\end{align}
One can easily recover the correct commutation relations $[S_{\ell j}^{'+},S_{\ell' j'}^{'-}]=2S^{'z}_{\ell j} \delta_{\ell\ell'}
\delta_{jj'}$ and $[S_{\ell j}^{'z}, S_{\ell' j'}^{'\pm}] =\pm S_{\ell j}^{'\pm} \delta_{\ell\ell'} \delta_{jj'}$.
Let us Fourier expand the creation and annihilation operators as
\begin{align}
  \widetilde c_{\ell j} = \frac{1}{\sqrt{N}} \sum_{\vec k \in \text{1BZ}} e^{-i\vec k\cdot \vec x_{\ell j}} c_{j, \vec k},~~~~~~
  \widetilde c_{\ell j}^\dagger = \frac{1}{\sqrt{N}} \sum_{\vec k \in \text{1BZ}} e^{i\vec k\cdot \vec x_{\ell j}} c_{j, \vec k}^\dagger,
  \label{FourierExpansion}
\end{align}
where $\vec x_{\ell j} = \vec x_\ell + \vec x_j$ is the position of the atomic cite labeled by $\ell$ and $j$ with $\vec x_\ell$ and $\vec x_j$ being the position of the center of the $\ell$-th magnetic unit cell and that of the $j$-th atom measured from the center, and
\begin{align}
	\left[ c_{j, \vec k}, c_{j', \vec k'}^\dagger \right] = \delta_{jj'} \delta_{\vec k,\vec k'}.
\end{align}
Hereafter, the summation of $\vec k$ is taken over the first Brillouin zone (1BZ) associated with magnetic unit cells.
Noting the equation
\begin{align}
  \sum_{\vec k} e^{i \vec k \cdot (\vec x_\ell - \vec x_{\ell'})} = N\delta_{\ell\ell'},
  ~~~~~~
	\sum_{\ell} e^{i(\vec k - \vec k')\cdot \vec x_{\ell}} =N \sum_{\vec G} \delta_{\vec k - \vec k', \vec G},
\end{align}
with the sum of the vector $\vec G$ is taken over all the reciprocal vectors,\footnote{
Note that the unique contribution to the calculation throughout this paper comes from $\vec G = 0$ since the sum over the magnon momentum covers only the first Brillouin zone.
}
the inverse transformation is given by
\begin{align}
	c_{j, \vec k} = \frac{1}{\sqrt{N}} \sum_\ell e^{i\vec k\cdot \vec x_{\ell j}} \widetilde c_{\ell j},~~~~~~
	c_{j, \vec k}^\dagger = \frac{1}{\sqrt{N}} \sum_\ell e^{-i\vec k\cdot \vec x_{\ell j}} \widetilde c_{\ell j}^\dagger.
\end{align}

Using the above relations, we can rewrite the Hamiltonian in a convenient form. Terms quadratic in $c_{j, \vec k}$ and $c_{j, \vec k}^\dagger$ represent the free Hamiltonian of the magnon, as soon shown below, and higher order terms represent its self interactions.
Note that, under the existence of non-zero matrix element $R_j^{z,x}, R_j^{z,y}$ or dipole interaction $D_{\ell\ell' jj'}^{\alpha \beta}$, there are terms of the form of $c_{j, \vec k} c_{j', \vec k'}$ and $c_{j, \vec k}^\dagger c_{j', \vec k'}^\dagger$ in the quadratic part of the Hamiltonian.
Thus we perform a Bogoliubov transformation to go to the canonical basis:
\begin{align}
	\begin{pmatrix} c_{j, \vec k} \\ c^\dagger_{j, -\vec k} \end{pmatrix} =
	\begin{pmatrix} u_{\vec k} & v_{\vec k} \\ v_{-\vec k}^* & u^*_{-\vec k} \end{pmatrix}
	\begin{pmatrix} \gamma_{\nu, \vec k} \\ \gamma^\dagger_{\nu, -\vec k} \end{pmatrix},
\end{align}
where $u_{\vec k} = \{ u_{j\nu, \vec k} \}$ and $v_{\vec k} = \{ v_{j\nu, \vec k} \}$ are $n \times n$ matrices with $\nu$ labeling $n$ different excitation modes.
By choosing proper matrices $u_{\vec k}$ and $v_{\vec k}$, we diagonalize the quadratic part of the Hamiltonian, which we denote by $H_0^{(\gamma)}$, as
\begin{align}
  H_0^{(\gamma)} = \sum_\nu \sum_{\vec k} \omega_{\nu, \vec k} \gamma_{\nu, \vec k}^\dagger \gamma_{\nu, \vec k}.
\end{align}
Thus $\gamma_{\nu, \vec k}$ and $\gamma_{\nu, \vec k}^\dagger$ represent the annihilation and creation operators of a quanta around the ground state, which is called magnon, and $\omega_{\nu, \vec k}$ denotes the dispersion relation of the magnon mode $\nu$.
In general, the magnon dispersion relation is anisotropic, i.e., $\omega_{\nu, \vec k}$ depends not only on $|\vec k|$ but also on the direction of $\vec k$~\cite{Herring:1951,Gurevich}.

As we will see later, only the lowest energy magnon mode around $k
\simeq 0$ is important for our discussion.  This mode, which is a
Nambu-Goldstone (NG) mode resulting from the symmetry breaking of the
spatial rotation, can be expressed in a much simpler effective
Hamiltonian.  We define the total spin operator $\vec S_\ell$ of the
$\ell$-th magnetic unit cell and the effective Hamiltonian
\begin{align}
  H_{\text{eff}} = -g \mu_B \sum_\ell \vec B^0 \cdot \vec S_\ell
  - \frac{J}{2} \sum_{\ell,\ell'} \vec S_\ell \cdot\vec S_{\ell'},
\end{align}
where the second sum is taken over the adjacent cells.  The above
effective Hamiltonian describes the NG mode as the unique magnon mode.
We can consider the Holstein-Primakoff transformation of the total
spin operator as
\begin{align}
  &S_{\ell}^{+} \equiv S_{\ell}^{x} + i S_{\ell}^{y}
  = \sqrt{2s} \sqrt{1-\frac{\widetilde c_{\ell}^\dagger \widetilde c_{\ell}}{2s}} \widetilde c_{\ell},\\
  &S_{\ell}^{-} \equiv S_{\ell}^{x} - i S_{\ell}^{y}
  = \sqrt{2s} \widetilde c_{\ell}^\dagger \sqrt{1-\frac{\widetilde c_{\ell}^\dagger \widetilde c_{\ell}}{2s}},\\
  &S_{\ell}^{z} = s - \widetilde c_{\ell}^\dagger \widetilde c_{\ell},
\end{align}
with
\begin{align}
  \left[ \widetilde c_\ell, \widetilde c_{\ell'}^\dagger \right] =
  \delta_{\ell\ell'}.
\end{align}
Here, $s$ is the size of the total spin of electrons inside a magnetic
unit cell.  With Fourier expanding $\widetilde c_\ell$ and $\widetilde
c_\ell^\dagger$ as Eq.\ \eqref{FourierExpansion}, we can see that
the quadratic part of $H_{\rm eff}$, which we call free Hamiltonian, is given by
\begin{align}
  H_{0} = \sum_{\vec k} \left[ \omega_L + 2Js \sum_p (1-\cos(\vec k \cdot \vec a_p)) \right] c_{\vec k}^\dagger c_{\vec k} \equiv \sum_{\vec k} \omega_{\vec k} c_{\vec k}^\dagger c_{\vec k},
  \label{H_magnon}
\end{align}
where $\omega_L \equiv g\mu_B B_z^0$ is the Larmor frequency with
$B_z^0$ being the $z$ component of the magnetic field $\vec B^0$, and
$\vec a_p$ $(p=1,2,3)$ are fundamental translation vectors that
generate magnetic unit cells.  For the YIG, we can use $s = 10$ and $J
= 0.35\,\text{meV}$, and the magnetic unit cell is a cube with $L
\equiv |\vec a_1| = |\vec a_2| = |\vec a_3| =
12.56\,\text{\AA}$~\cite{Cherepanov:1993}.

Let us focus on the material with the cubic unit cell for simplicity.
In the long wavelength limit $|\vec k| L \ll 1$, the dispersion relation is given by
\begin{align}
	\omega_{\vec k} \simeq \omega_L + JsL^2 k^2 \equiv \omega_L + \frac{k^2}{2M},
\end{align}
with $k \equiv |\vec k|$.
Note that this mode is the so-called type II Nambu-Goldstone boson according to the classification proposed in \cite{Watanabe:2012hr,Hidaka:2012ym} and thus $\omega_{\vec{k}} \propto \vec{k}^2$ when $B_z^0 = 0$.
The ferromagneticity of the material is responsible for this classification; the commutator of generators of two broken symmetries, e.g., the rotation around $x$ and $y$ axes in the global coordinate, is proportional to the angular momentum operator along the $z$ axis, which possesses a non-zero expectation value at the ground state.
This means that two broken generators are dependent on each other, which results in only one type II Nambu-Goldstone boson.

The $k=0$ mode corresponds to the homogeneously rotating mode around the external magnetic field with Larmor frequency, which is called the Kittel mode. In a typical material, $M \sim \mathcal O(1)$\,MeV; for example, using the values shown above, we obtain $M \sim 3.5\,\text{MeV}$ for the YIG. The Larmor frequency is evaluated as
\begin{align}
	\omega_L = \frac{eB_z^0}{m_e} \simeq 1.2\times 10^{-4}\,{\rm eV}\left( \frac{B_z^0}{1\,{\rm T}} \right).
\end{align}
For the purpose of DM detection discussed below, the DM detection rate is enhanced if the Larmor frequency is close to the DM mass, and hence we are interested in the DM mass of meV range.\footnote{
	Ref.~\cite{Trickle:2019ovy} considered DM scattering with an electron as an excitation process of magnon. It may be interpreted as the magnon emission by DM. On the other hand, we consider DM absorption by the electron, which may be regarded as the DM conversion into a magnon. In the latter case, it is essential to apply the magnetic field to control the gap of the magnon dispersion relation.
}

\section{Axion conversion into magnon}
\label{sec:conv}

First, we consider the case of axion DM which interacts with the
electron and calculate the axion-magnon conversion rate.
In Ref.~\cite{Barbieri:2016vwg}, a classical calculation was used to
estimate the axion-magnon conversion rate.
We take a quantum mechanical method to calculate the conversion rate and show
that it reproduces the result of Ref.~\cite{Barbieri:2016vwg}.
A quantum mechanical calculation of the conversion rate with a slightly different manner has been done in Ref.~\cite{Flower:2018qgb}
and the result is also consistent with ours.
An advantage of the quantum mechanical calculation is that it is
applicable even in the case where only a small number of magnons are
excited during the time scale of our interest.  We then apply the same
method to the hidden photon DM.

\subsection{Formulation}

The axion (denoted by $a$) is assumed to interact with the electron,
as in the DFSZ model~\cite{Zhitnitsky:1980tq,Dine:1981rt} or the
flaxion/axiflavon~\cite{Ema:2016ops,Calibbi:2016hwq}. The Lagrangian
density is
\begin{align}
	\mathcal L = \frac{1}{2}(\partial_\mu a)^2-\frac{1}{2}m_a^2a^2 + \overline\psi (i\Slash{\partial} - m_e) \psi + \frac{\partial_\mu a}{2f} \overline\psi \gamma^\mu\gamma_5 \psi,
\label{L_DFSZ}
\end{align}
where $\psi$ denotes the electron and $f$ is of the order of the
Peccei-Quinn symmetry breaking scale.  Then, in the non-relativistic
limit of the electron, the total interaction Hamiltonian of the
material is
\begin{align}
  H_{\rm int} = \frac{1}{f} \sum_\ell \vec\nabla a(\vec x_\ell)\cdot \vec S_\ell,
\end{align}
where $\vec S_\ell$ is the electron spin at each cite $\ell$ (see
Appendix \ref{sec:heff}).

Below we treat the axion as a classical background described by
\begin{align}
  a(\vec x,t) = a_0 \cos(m_a t - m_a \vec{v}_a \cdot\vec x + \delta),
\end{align}
with $v_a\ll 1$.  This treatment is valid within the axion coherence
time $\tau_a \sim (m_a v_a^2)^{-1}$.  Note that $m_a a_0 =
\sqrt{2\rho_{\rm DM}}$, with $\rho_{\rm DM} \sim 0.3\,\mathrm{GeV}/\mathrm{cm}^3$ being the energy density of DM.
In the following, the location of the ferromagnetic material is chosen
to be close to the origin $\vec{x}\sim 0$.  Then, the interaction Hamiltonian becomes
\begin{align}
  H_{\rm int} =
  \frac{m_a a_0 v_a}{f} \sum_\ell
  \vec e_v\cdot \vec S_\ell\, \sin(m_a t + \delta),
  \label{Hint_axion}
\end{align}
where $\vec e_v$ is the unit vector pointing to the direction of $\vec v_a$.
At the first order in the magnon creation or
annihilation operator, we obtain
\begin{align}
  H_{\rm int} = \frac{m_a a_0 \sin(m_a t+ \delta)}{f} \sqrt{\frac{s}{2}} \sum_\ell \left( v_a^- \widetilde c_\ell + v_a^+ \widetilde c_\ell^\dagger \right)
	= \sin(m_a t+ \delta) \left(V^* c_0 + V c_0^\dagger \right),
\end{align}
where we used the fact that $(m_av_a)^{-1}$ is expected to be much larger
than the size of the ferromagnetic material.
In addition, we define
\begin{align}
	v_a^\pm \equiv v^x_a \pm i v^y_a,  ~~~~~~~~V\equiv \sqrt{\frac{sN}{2}}\frac{m_a a_0 v_a^+}{f},
\end{align}
with choosing the direction of $\vec S_\ell$ in the ground state as the $z$-axis.
Note that only the $\vec{k}=0$ magnon mode contributes to $H_{\mathrm{int}}$ evaluated at the first order because of the approximately homogeneous nature of the axion background compared with the material size.
The total magnon-axion Hamiltonian is
\begin{align}
	H =H_0 + H_{\rm int},
\end{align}
where the magnon free Hamiltonian $H_0$ is given in Eq.\ (\ref{H_magnon}).

Now let us estimate the axion-magnon conversion rate based on the Hamiltonian derived above.
For the axion-magnon conversion, only the $k\simeq 0$ mode matters since the axion momentum is negligible compared with its mass. The magnon has a dispersion relation $\omega_k = \omega_L + k^2/(2M)$ and $\omega_L$ is chosen such that $\omega_L \simeq m_a$.
The system can be approximated by a two-level system: the ground state $\left|0\right>$ and the excited state $\left|1\right>$ which is defined by $c_0^\dagger \left|0\right>$. In principle, there are higher excited states $\left(c_0^\dagger \right)^n \left| 0\right>$ $(n\geq 2)$, but the probability to reach to these states is negligibly small for the situation of our interest.
The quantum state $\left|\psi(t)\right>$ is, in general, a linear superposition of them:
\begin{align}
	\left|\psi(t)\right> = \alpha_0(t)  \left|0\right> + \alpha_1(t) \left|1\right>.
\end{align}
The initial condition is taken to be $\alpha_0(t=0)=1$ and $\alpha_1(t=0)=0$. The Schrodinger equation is
\begin{align}
	i \frac{\partial}{\partial t}\left|\psi(t)\right> = (H_0 + H_{\rm int}) \left|\psi(t)\right>.
\end{align}
It is convenient to go to the interaction picture: let us define $\left|\phi(t)\right>\equiv  e^{iH_0 t}\left|\psi(t)\right>$. Then the Schrodinger equation becomes
\begin{align}
	i \frac{\partial}{\partial t}\left|\phi(t)\right> = e^{iH_0 t}H_{\rm int} e^{-i H_0 t} \left|\phi(t)\right>.
\end{align}
From this, we obtain the differential equation
\begin{align}
	&i \dot\alpha_0 = V^* \sin(m_at+ \delta) \alpha_1,\\
	&i\dot \alpha_1 = \omega_L \alpha_1 + V \sin(m_at+ \delta) \alpha_0.
\end{align}
Assuming $|V| \ll \omega_L$, which is valid in parameters of our interest, it is solved as
\begin{align}
	\alpha_1(t) \simeq \frac{iV}{2}\frac{ e^{i\delta}(m_a-\omega_L)(e^{im_at}-e^{-i\omega_Lt}) + e^{-i\delta}(m_a+\omega_L)(e^{-im_at}-e^{-i\omega_L t}) }{m_a^2-\omega_L^2}.
\end{align}
The probability that we find the state $\left|1\right>$ at the time $t$ is given by $P(t)=|\alpha_1(t)|^2$.
Clearly, the probability is enhanced for $\omega_L \simeq m_a$. In this case, we have
\begin{align}
	P(t) \simeq \frac{|V|^2 t^2}{4}.
\end{align}

The excited magnon is detected through its coupling to the cavity photon.
In the QUAX setup, the cavity photon mode is chosen such that the cavity frequency $\omega_{\rm cav}$ coincides with $\omega_L$. In this case, the hybridization (or the mixing) between cavity and Kittel mode takes place, and the magnon should be regarded rather as a polariton (or ``magnon-polariton'')~\cite{Zhang:2014,Tabuchi:2014,Tabuchi:2015}.
Including the cavity mode and focusing only on the zero mode, the Hamiltonian is given by
\begin{align}
  H &= \omega_L c_0^\dagger c_0 + \omega_{\rm cav} b^\dagger b + g_{\rm cm} (b^\dagger c_0 + c_0^\dagger b) \\
  &= (\omega_L+g_{\rm cm}) c_+^\dagger c_+ + (\omega_L-g_{\rm cm})c_-^\dagger c_-,
\end{align}
where $b^\dagger$ ($b$) is the creation (annihilation) operator of the cavity mode, $g_{\rm cm}$ represents the cavity-magnon coupling rate,\footnote{
	The photon-magnon mixing comes from the dipole interaction $H=-g\mu_B \sum_\ell \vec B(\vec x_\ell)\cdot \vec S_\ell$. The mixing parameter is roughly given by $g_{\rm cm} \sim g\mu_B \sqrt{2sN} V_{\rm cav}^{-2/3}$ where $V_{\rm cav}$ is the cavity volume. In the QUAX setup $\omega_L\gg g_{\rm cm}$.
}$c_{\pm}\equiv (c_0\pm b)/\sqrt{2}$ and we have taken $\omega_L=\omega_{\rm cav}$ in the last line.
Thus two modes are maximally mixed and all the energy eigenstates are generated by one-to-one superposition of $b^\dagger$ and $c_0^\dagger$.
Accordingly, when many magnon modes are excited, half of them are detected as a cavity mode after their propagation.
Thus the power obtained by the transition is given by
\begin{align}
  \frac{dE_{\rm signal}}{dt} = \frac{\omega_L P(t)}{2t} = \frac{\omega_L |V|^2 t}{8}.  \label{power_magnon}
\end{align}
It is consistent with classical calculation in \cite{Barbieri:2016vwg}
(see also Appendix \ref{sec:classical}). Note that $t$ is limited by
the axion coherence time $\tau_a$ or the magnon-polariton relaxation time $\tau_m$ (due to spin-lattice and spin-spin interactions and dissipation of cavity mode),
whichever is smaller determines the effective coherence time through $\tau \equiv {\rm min}[\tau_a, \tau_m]$.
The event rate is then
\begin{align}
  \left[\frac{dN_{\rm signal}}{dt}\right]_{\rm spin} &= \frac{P(\tau)}{2\tau} = \frac{|V|^2 \tau}{8}=\frac{sN}{4} \frac{\rho_{\rm DM}(v_a^{x2}+v_a^{y2})\tau}{f^2}.
\end{align}
To derive more convenient expression, we convert the factor $s N$ to the target mass $M_{\rm target}$ through
\begin{align}
  M(T) M_{\rm target} = g \frac{e}{2m} s N,
\end{align}
where $M(T)$ is the magnetization of the target.
Hereafter, we assume the target material to be YIG at temperature $T \sim 100\,\mathrm{mK}$ according to the QUAX proposal, which yields $M \simeq 38\, \mathrm{emu / g}$ \cite{Cherepanov:1993}.
Substituting all the above, we obtain
\begin{align}
 \left[ \frac{dN_{\rm signal}}{dt}\right]_{\rm spin}
  \simeq 0.05\,{\rm s^{-1}}\left( \frac{M_{\rm target}}{1\,{\rm kg}} \right)
  \left( \frac{10^{10}\,{\rm GeV}}{f} \right)^2 \left( \frac{\tau}{2\,{\rm \mu s}} \right)\left( \frac{v_a}{10^{-3}} \right)^2 \sin^2\theta,
\end{align}
where $\theta$ is the angle between $\vec v_a$ and $z$ direction.

\subsection{Sensitivity}  \label{sec:sens_axion}

So far we have discussed the axion-spin interaction.
One should note that the cavity setup also works as a standard haloscope~\cite{Sikivie:1983ip,Bradley:2003kg} if the axion has a Chern-Simons coupling like
\begin{align}
	\mathcal L = -C_{a\gamma}\frac{\alpha_e}{8\pi} \frac{a}{f} F_{\mu\nu} \widetilde F^{\mu\nu}
	= C_{a\gamma}\frac{\alpha_e}{2\pi} \frac{a}{f} \vec B\cdot \vec E,
	\label{LTI_axion}
\end{align}
where $\widetilde F^{\mu\nu} \equiv \epsilon^{\mu\nu\rho\sigma}F_{\rho\sigma}/2$, $\alpha_e$ is the electromagnetic fine structure constant and $C_{a\gamma}$ is an $\mathcal O(1)$ model-dependent coupling constant.
The background DM axion generates the cavity mode under the applied magnetic field.
The photon event rate is estimated as~\cite{Sikivie:1983ip,Bradley:2003kg}
\begin{align}
	\left[\frac{dN_{\rm signal}}{dt}\right]_{\rm CS} &= \left( \frac{C_{a\gamma} \alpha_e}{2\pi f} \right)^2 \frac{\rho_{\rm DM} B_0^2}{m_a} V_{\rm cav}\mathcal G_{\rm cav}{\rm min}\left[\tau_a,\tau_{\rm cav}\right] \\
	&\simeq 7.1\times 10^{-1}\,{\rm s^{-1}} C_{a\gamma}^2
	\left( \frac{10^{-4}\,{\rm eV}}{m_a} \right)\left( \frac{10^{10}\,{\rm GeV}}{f} \right)^2\left( \frac{B_0}{1\,{\rm T}} \right)^2
	\left( \frac{V_{\rm cav} \mathcal G_{\rm cav}}{100\,{\rm cm^3}} \right)\left( \frac{\tau_{\rm cav}}{2\,{\rm \mu s}} \right),
\end{align}
where $\mathcal G_{\rm cav}$ is an $\mathcal O(1)$ form factor which depends on a cavity mode, $V_{\rm cav}$ is the cavity volume and $\tau_{\rm cav}$ is the cavity decay time.\footnote{
	Generally speaking, $\tau_{\rm cav}$ can be different from $\tau_m$ since the latter includes the effect of spin relaxation time. For simplicity, however, we take $\tau_m \simeq \tau_{\rm cav}$ as assumed in Ref.~\cite{Barbieri:2016vwg}.
}
Comparing it with $\left[dN_{\rm signal} /dt \right]_{\rm spin}$, the signal induced by Chern-Simons coupling may not be neglected.
The relative ratio of these two signals depends on the target mass of the ferromagnet and cavity volume.
Note that the axion-spin coupling is greatly suppressed in the KSVZ axion model~\cite{Kim:1979if,Shifman:1979if}, and hence in such a model only the signal from the Chern-Simons coupling is relevant.
Thus one can distinguish the axion model by comparing the signal with and without insertion of the ferromagnetic material inside the cavity.

Let us estimate the experimental sensitivity following Ref.~\cite{Barbieri:2016vwg}.
We evaluate the sensitivity for both an ideal setup using a single photon counter \cite{Lamoreaux:2013koa, Capparelli:2015mxa} and a more realistic setup using a linear amplifier.
For an ideal setup, we consider only thermal fluctuation as a source of the noise.
The noise rate is given by
\begin{align}
	\frac{dN_{\rm noise}}{dt} \sim \frac{1/\tau_{\rm cav}}{\exp(m_a/T_{\rm cav})-1},
  \label{eq:noise_rate}
\end{align}
where $T_{\rm cav}$ is the cavity temperature. For example, for $T_{\rm cav}=116$\,mK, the noise rate is about $dN_{\rm noise}/dt \simeq 10^{-3}$\,Hz at $m_a=200$\,$\mathrm{\mu eV}$ and $\tau_{\mathrm{cav}}=2\,\mathrm{\mu s}$. The signal-to-noise ratio (SNR) during the observation time $T_{\rm obs}$ for each scan is given by
\begin{align}
  {\rm SNR} = \frac{(dN_{\rm signal}/dt) T_{\rm obs}}{\sqrt{(dN_{\rm noise}/dt)  T_{\rm obs} }}.
\end{align}
Requiring ${\rm SNR} \sim (\text{a few})$ (below, we use $\text{SNR} = 3$ to evaluate the sensitivity), one obtains a minimal observation time $T_{\rm obs}$ for each DM parameter.
On the other hand, the bandwidth of the magnon-polariton is about $\Delta \omega \sim 1$\,MHz, while the effective coherence time is given by $\tau \simeq \tau_m = 2/\Delta \omega \simeq 2\,\mu$s at the target axion mass $m_a = 200\,\mu$eV. Thus the covered DM mass range during the total observation time $T_{\rm total}$ (say, $T_{\rm total} \sim 10$ years) is given by $\Delta m_a \simeq \Delta \omega\times (T_{\rm total}/T_{\rm obs})$.

On the other hand, for a realistic setup using a linear amplifier, the size of the noise and the observation time $T_{\mathrm{obs}}$ determines the minimal measureable power $P_{\mathrm{min}}$.
According to the Dicke radiometer equation,
\begin{align}
  P_{\mathrm{min}} = T_{\mathrm{noise}} \sqrt{\frac{\Delta\omega}{T_{\mathrm{obs}}}},
\end{align}
where $T_{\mathrm{noise}}$ is the noise temperature that characterizes the size of noise in this observation.
To evaluate $P_{\mathrm{min}}$, we assume that the quantum noise dominates all the other sources of noise and use $T_{\mathrm{noise}} = \omega_L$.\footnote{
For this assumption to be true, we should at least prepare a sufficiently low temperature cavity with $T_c \ll m_a$ to suppress the thermal noise.
}
If the expected output power from axion $\omega_L (d N_{\mathrm{signal}} / d t)$ is larger than $P_{\mathrm{min}}$, we can detect the effect from axion using this setup.

\begin{figure}[t]
  \includegraphics[width=0.48\hsize]{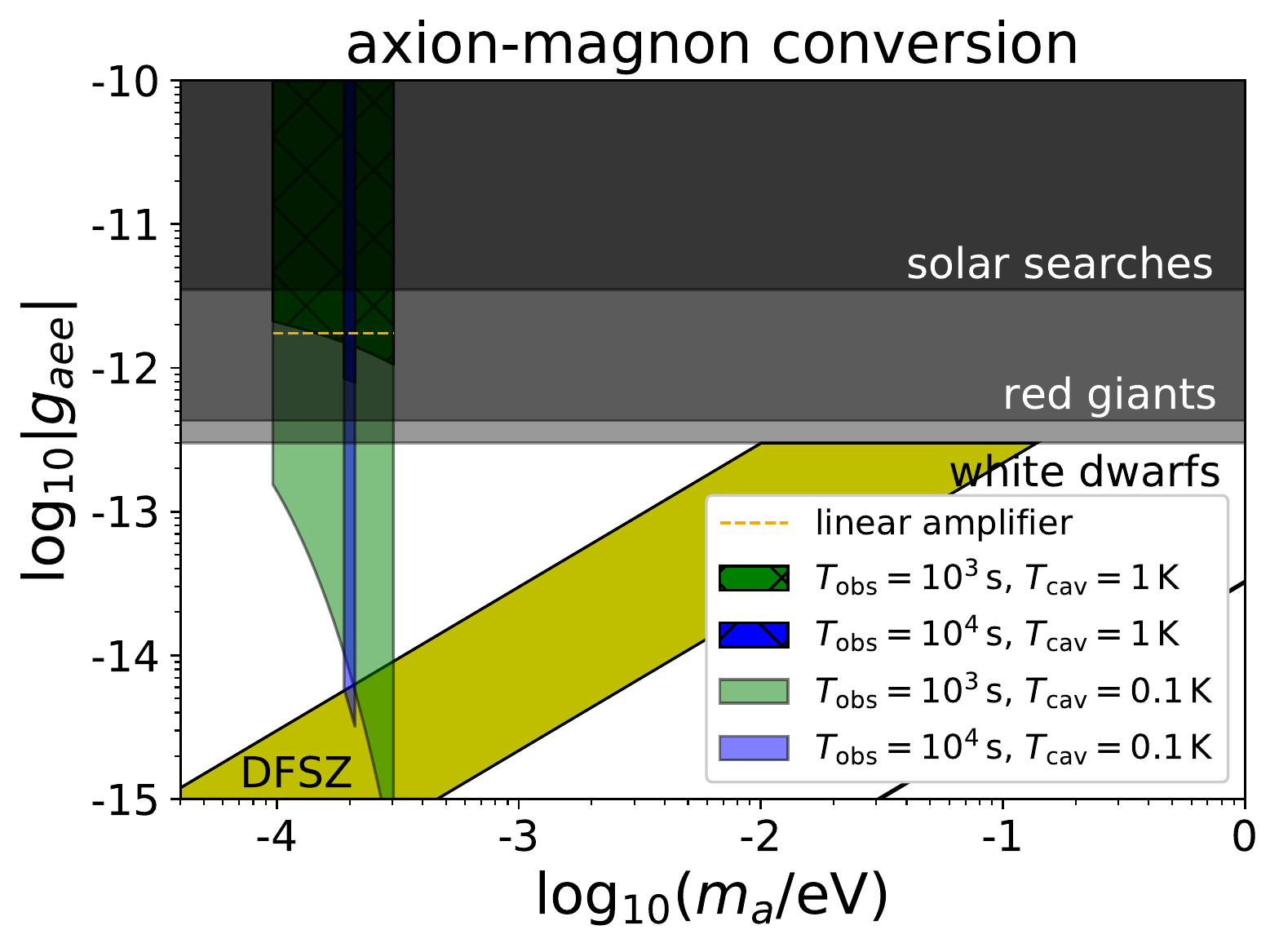}
  \includegraphics[width=0.48\hsize]{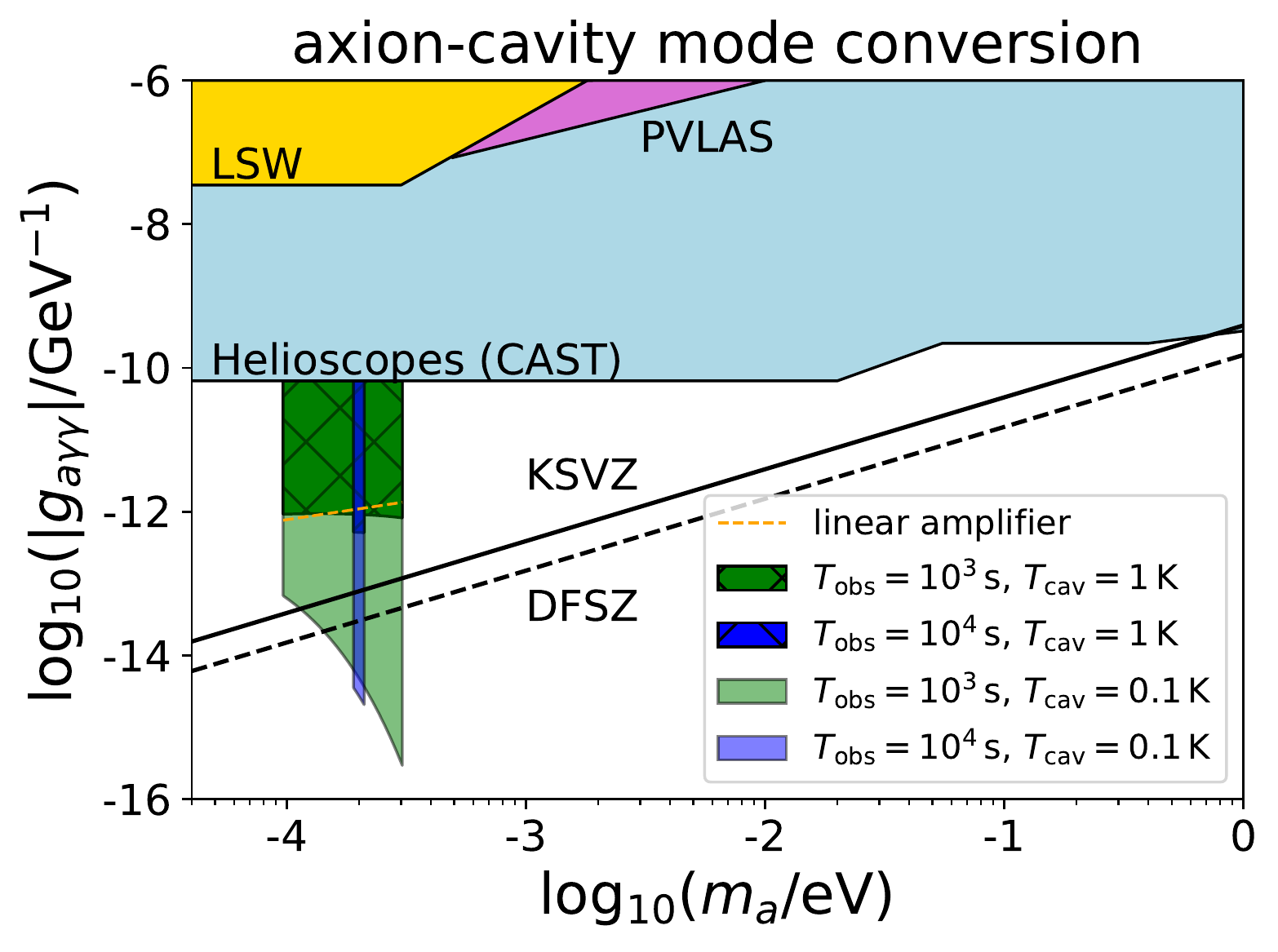}
  \caption{
    Sensitivity plot for $\text{SNR} = 3$ under $T_{\text{total}} = 10\,\text{years}$.
    \textit{Left}: Sensitivity of the magnon detector on the axion-electron coupling $g_{aee}$ as a function of the axion mass $m_a$.
    The green and blue regions show the sensitivity for an ideal setup.
    The colors and styles of regions represent different setups; the observation time for each scan is set to be $T_{\mathrm{obs}} = 10^3\,\mathrm{s}$ (green) or $10^4\,\mathrm{s}$ (blue), and the cavity temperature is $T_{\mathrm{cav}} = 1\,\mathrm{K}$ (dark-meshed) or $0.1\,\mathrm{K}$ (light).
    The orange dashed line shows the sensitivity for a realistic setup with $T_{\mathrm{obs}}=10^3\,\mathrm{s}$ and $T_{\mathrm{cav}} \ll m_a$.
    Throughout the figure, the setup of $M_{\text{target}} = 1\,\mathrm{kg}$, $\tau=2\,\mathrm{\mu s}$, $v_a = 10^{-3}$, and $\sin^2\theta = 0.5$ is assumed.
    Besides, the gray regions show the parameter region already excluded by other searches and the yellow region and the black solid line correspond to the prediction of the DFSZ model with $0.28 \lesssim \tan\beta \lesssim 140$ and that of the KSVZ model, respectively.
    \textit{Right}: Sensitivity of the cavity detector on the axion-photon coupling $g_{a\gamma\gamma}$ as a function of $m_a$.
    Similar to the left panel, the green and blue regions and orange lines show the sensitivities with $B_0 = 1\,\mathrm{T}$, $V_{\text{cav}} \mathcal{G}_{\text{cav}} = 100\,\mathrm{cm}^3$, and $\tau_{\text{cav}} = 2\,\mathrm{\mu s}$.
    The other shaded regions show the region excluded by other searches and the black dashed (solid) line corresponds to the prediction of the DFSZ (KSVZ) model.
  }
  \label{fig:axion}
\end{figure}

In Fig.~\ref{fig:axion}, we show the sensitivity of the magnon detector (left) and the cavity detector (right) with the total observation of $T_{\text{total}} = 10\,\text{years}$.
For an ideal setup, we use two different choices of the observation time $T_{\text{obs}} = 10^3\,\mathrm{s}$ and $10^4\,\mathrm{s}$, which are shown by green and blue colors, respectively.
We also use two different choices of the cavity temperature $T_{\mathrm{cav}} = 1\,\mathrm{K}$ and $0.1\,\mathrm{K}$, which are shown by the dark-meshed and the light regions, respectively.
For a realistic setup, we use the choice of $T_{\mathrm{obs}} = 10^3\,\mathrm{s}$ and $T_{\mathrm{cav}} \ll m_a$ and the result is shown with an orange dashed line.
The center of the scanned region of the axion mass is fixed to be $m_a = 200\,\mathrm{\mu eV}$ and the width of the region is given by $\Delta m_a$ for each choice of $T_{\text{obs}}$.

In the left panel, the sensitivity on the dimensionless axion-electron
coupling $g_{aee} \equiv m_e / f$ is shown as a function of the axion
mass $m_a$, assuming the setup of $M_{\text{target}} =
1\,\mathrm{kg}$, $\tau=2\,\mathrm{\mu s}$, $v_a = 10^{-3}$, and
$\sin^2\theta = 0.5$.
Gray regions correspond to the parameter space
excluded by other searches using the bremsstrahlung from white dwarfs
\cite{Bertolami:2014wua}, the brightness of the tip of the red-giant
branch in globular clusters~\cite{Viaux:2013lha}, and the direct
detection of solar axions at the
EDELWEISS-II~\cite{Armengaud:2013rta}, the
XENON100~\cite{Aprile:2014eoa}, and the LUX~\cite{Akerib:2017uem}
collaborations.  Besides, the yellow region and the black solid line
show the prediction for the DFSZ and KSVZ models, respectively.  To
obtain the DFSZ prediction, we variate $\tan\beta$, which is the ratio
between vacuum expectation values of the two Higgs doublets, within
$0.28 \lesssim \tan\beta \lesssim 140$ as required by the perturbative
unitarity of Yukawa couplings~\cite{Chen:2013kt}.  By comparing with the right panel, we can see that the
axion search using the cavity mode has a better sensitivity than that
using magnon excitation for the DFSZ and KSVZ models.
At the same time, however, the sensitivity of the magnon detector reaches the DFSZ prediction for a relatively heavy mass due to the Boltzmann suppression of the noise rate according to Eq.~\eqref{eq:noise_rate}.
Thus, the figure shows the potential to probe the axion-electron coupling depending on the details of the model.
It opens up a possibility to distinguish the KSVZ and DFSZ model by looking at the magnon-induced signal, once the DM signal is discovered at the cavity experiment.

In the right panel, the sensitivity on the axion-photon coupling $g_{a\gamma\gamma} \equiv C_{a\gamma} \alpha_e / (2\pi f)$ is shown, assuming the setup of $B_0 = 1\,\mathrm{T}$, $V_{\text{cav}} \mathcal{G}_{\text{cav}} = 100\,\mathrm{cm}^3$, and $\tau_{\text{cav}} = 2\,\mathrm{\mu s}$ in this case.
Three shaded regions in $|g_{a\gamma\gamma}| \gtrsim 10^{-10}\,\mathrm{GeV}^{-1}$ correspond to the regions excluded by existing searches; the helioscope CAST~\cite{Anastassopoulos:2017ftl}, the Light-Shining-through-Walls (LSW) experiments such as the OSQAR~\cite{Ballou:2015cka}, and the measurement of the vacuum magnetic birefringence at the PVLAS~\cite{DellaValle:2015xxa}.
The black dashed (solid) line corresponds to the prediction of the DFSZ (KSVZ) model.
We can see that the predicted values of $g_{a\gamma\gamma}$ around $m_a\simeq 200\,\mathrm{\mu eV}$ are covered by the cavity detector with the given setup.

\section{Hidden photon conversion into magnon}  \label{sec:hidden}

\subsection{Formulation}

Let us consider a model with a massive hidden photon which has a
kinetic mixing with hypercharge photon.  In such a model, the hidden
photon interacts with SM fields via
\begin{align}
  \mathcal L = -\frac{1}{4} H_{\mu\nu}H^{\mu\nu} - \frac{1}{4} B_{\mu\nu}B^{\mu\nu} + \frac{\epsilon_Y}{2} H_{\mu\nu}B^{\mu\nu}
  +\frac{1}{2} m_H^2 H_\mu H^\mu,
\label{L_hiddenphoton}
\end{align}
where $H_{\mu\nu}=\partial_\mu H_\nu-\partial_\nu H_\mu$ with $H_\mu$
being the hidden photon, $B_{\mu\nu}=\partial_\mu B_\nu-\partial_\nu
B_\mu$ with $B_\mu$ being the hypercharge photon and $\Phi$ denotes
the SM Higgs doublet.  The kinetic terms of gauge bosons are
diagonalized by the following transformation:
\begin{align}
	B_\mu' = B_\mu - \epsilon_Y H_\mu,~~~~~~H_\mu'=\sqrt{1-\epsilon_Y^2} H_\mu,
\end{align}
so that the kinetic term of the hidden photon and hypercharge photon and the hidden photon mass term become
\begin{align}
	\mathcal L = -\frac{1}{4} H'_{\mu\nu}H^{\prime\mu\nu} - \frac{1}{4} B'_{\mu\nu}B^{\prime\mu\nu} + \frac{1}{2} m_H^{\prime 2} H'_\mu
H^{\prime\mu},
\end{align}
where $m_H^{\prime 2} = m_H^2/(1-\epsilon_Y^2)$. After the Higgs
obtains a VEV, these gauge bosons, as well as the neutral weak gauge
boson $(W^3_\mu)$, are mixed. Denoting the Higgs VEV as $v\simeq 174\,$GeV, the
gauge boson mass term is given by
\begin{align}
	\mathcal L_{\rm mass} = \frac{m_Z^2}{2}\left( c_W W_\mu^3- s_W B_\mu \right)^2 + \frac{m_H^{\prime 2}}{2} H_\mu^{\prime 2}.
\end{align}
Here, $m_Z^2=(g_W^2+g_Y^2)v^2/2$, $c_W=g_W/\sqrt{g_W^2+g_Y^2}$, where
$s_W=g_Y/\sqrt{g_W^2+g_Y^2}$, $g_W$ is the weak gauge coupling and
$g_Y$ is the hypercharge gauge coupling $(g_Y=e/c_W)$.  Besides,
the SM fermions (denoted as $\psi$) are neutral for the hidden photon gauge interaction and hence the interactions between the SM fermions
and hidden photon originate from
\begin{align}
  \mathcal L_{\rm int} =
  \overline{\psi}
  (Q_Y g_Y \gamma^\mu B_\mu + Q_{T_3}g_W \gamma^\mu W^3_\mu)
  \psi,
\end{align}
where $Q_Y$ is the hypercharge of $\psi$ and $Q_{T_3}$ is the charge
under $T_3$ rotation of SU(2).

The mass matrix in the $(W_\mu^3,B_\mu', H_\mu')$ basis is
\begin{align}
	\mathcal M^2= m_Z^2\begin{pmatrix}
		c_W^2 & -s_Wc_W & s_W c_W \epsilon_Y/\sqrt{1-\epsilon_Y^2} \\
		-s_Wc_W & s_W^2 & -s_W^2 \epsilon_Y/\sqrt{1-\epsilon_Y^2} \\
		s_W c_W \epsilon_Y/\sqrt{1-\epsilon_Y^2} & -s_W^2\epsilon_Y/\sqrt{1-\epsilon_Y^2} & s_W^2 \epsilon_Y^2/(1-\epsilon_Y^2)+m_H^{\prime 2}/m_Z^2
	\end{pmatrix}.
\end{align}
Up to the first order in $\epsilon_Y$, It is diagonalized by the following unitary transformation to go to the mass eigenstate $(Z_\mu, A_\mu, H_\mu')$:
\begin{align}
	\begin{pmatrix}
		W_\mu^3 \\ B_\mu' \\ H_\mu'
	\end{pmatrix}
	\simeq
	\begin{pmatrix}
		c_W & s_W & -s_W c_W \epsilon_Y \\
		-s_W & c_W & s_W^2 \epsilon_Y \\
		s_W \epsilon_Y & 0 & 1
	\end{pmatrix}
	\begin{pmatrix}
		Z_\mu \\ A_\mu \\ H_\mu''
	\end{pmatrix},
\end{align}
with the mass eigenvalues of $(m_Z^2, 0, m_H^{\prime 2})$.

After integrating out $Z$-boson, whose mass is much larger than the
energy scale of our interest, the relevant part of the fermion
interaction can be written as
\begin{align}
  \mathcal L_{\rm int} \supset
  - \epsilon \,e Q \,H_\mu'' \,\overline \psi \gamma^\mu \psi,
\end{align}
where $Q = Q_Y + Q_{T_3}$ is the electromagnetic charge and $\epsilon
\equiv \epsilon_Y c_W$.  Thus, the hidden photon effectively couples
to electromagnetic current with an effective coupling constant $\epsilon
e$.  The interaction Hamiltonian with electron in the non-relativistic
limit is then given by
\begin{align}
  H_{\rm int} = - \frac{\epsilon eQ}{m_e}
  \sum_\ell \vec{B}_H (\vec{x}_\ell) \cdot \vec{S}_\ell,
\end{align}
where $\vec{B}_H\equiv \vec{\nabla}\times\vec{H}$ is the hidden
magnetic field (see Appendix \ref{sec:heff}).

We parametrize the hidden photon background as
\begin{align}
  H_0(t,\vec x) = &\,
  -\vec v_H \cdot \vec{\widetilde H}\cos\left(m_H t - m_H \vec{v}_H \cdot\vec x + \delta \right), \\
  \vec H(t,\vec x) = &\, \vec{\widetilde H}\cos\left(m_H t - m_H \vec{v}_H \cdot\vec x + \delta \right),
\end{align}
to satisfy the equation of motion $(\square + m_H^2) \vec H = 0$ and
the Lorentz condition $\partial_\mu H^\mu=0$.  At the location of the
ferromagnetic material, the hidden electric and magnetic fields are
given by
\begin{align}
  \vec E_H \simeq &\,
  \vec{\widetilde H}\,m_H  \sin\left(m_H t + \delta \right),\\
  \vec B_H \simeq &\,
  \vec v_H\times \vec{\widetilde H}\,m_H \sin\left(m_H t + \delta \right).
\end{align}
The DM density is given by $\rho_{\rm DM} = m_H^2 \widetilde H^2/2$.

The hidden photon-magnon interaction Hamiltonian is written as
\begin{align}
  H_{\rm int} =
  \frac{\epsilon e m_H \widetilde H v_H}{m_e}
  \sum_\ell
  \vec e_{B} \cdot \vec S_\ell \sin\varphi\,\sin(m_H t + \delta),
  \label{Hint_HP_magnon}
\end{align}
where $\varphi$ denotes the angle between $\vec v_H$ and $\vec H$, and
$\vec e_{B}$ is the unit vector of the direction of $\vec B_H$.  It
causes hidden photon-magnon conversion under the static magnetic field
as in the case of the axion. Comparing (\ref{Hint_HP_magnon}) with the
axion-magnon Hamiltonian (\ref{Hint_axion}), we can repeat the same
analysis by just reinterpreting $1/f \to \epsilon e\sin\varphi/m_e$.
Thus, referring to (\ref{power_magnon}), the power obtained by this
process is given by
\begin{align}
  \frac{dE_{\rm signal}}{dt} = \frac{(\epsilon e)^2 \omega_L sN \rho_{\rm DM} v_H^2 t}{8 m_e^2}\sin^2\theta \sin^2\varphi,
\end{align}
where $\theta$ denotes the angle between $\vec B_H$ and $z$-axis.
Note again that $t$ is limited by the hidden photon coherence time or
magnon relaxation time (due to spin-lattice or spin-spin
interactions), which we denote by $\tau$. The event rate is then
\begin{align}
  \left[\frac{dN_{\rm signal}}{dt}\right]_{\rm spin} &= \frac{(\epsilon e)^2 sN \rho_{\rm DM} v_H^2 \tau}{8 m_e^2}\sin^2\theta \sin^2\varphi
  \nonumber \\
  &\simeq 9.6\times 10^{-5}\,{\rm s^{-1}}\left( \frac{\epsilon}{10^{-14}}\right)^2\left( \frac{M_{\text{target}}}{1\,\text{kg}} \right)
  \left( \frac{v_H}{10^{-3}} \right)^2 \left( \frac{\tau}{2\,{\rm \mu s}} \right) \sin^2\theta \sin^2\varphi,
\end{align}
where we use the same setup like that in the previous section to convert $sN$ into $M_{\text{target}}$.

\subsection{Sensitivity}

So far we have discussed the hidden photon interaction with electron spin and its consequences for magnon excitation.
However, as in the case of axion DM, the cavity setup itself also works as a hidden photon detector even without magnetic material~\cite{Wagner:2010mi,Arias:2012az}.
The background DM hidden photon generates the cavity mode through the kinetic mixing term and the photon event rate is estimated as~\cite{Arias:2012az}
\begin{align}
	\left[\frac{dN_{\rm signal}}{dt}\right]_{\rm mix} &=  \epsilon^2 m_H \rho_{\rm DM}V_{\rm cav}\mathcal G_{\rm cav}{\rm min}\left[\tau_H,\tau_{\rm cav}\right] \\
	&\simeq 1.4\times 10^{3}\,{\rm s^{-1}}
	\left( \frac{\epsilon}{10^{-14}} \right)^2\left( \frac{m_H}{10^{-4}\,{\rm eV}} \right)
	\left( \frac{V_{\rm cav} \mathcal G_{\rm cav}}{100\,{\rm cm^3}} \right)\left( \frac{\tau_{\rm cav}}{2\,{\rm \mu s}} \right),
\end{align}
where $\mathcal G_{\rm cav}$ is an $\mathcal O(1)$ form factor which may take a different value from the axion DM case.
Comparing it with $\left[dN_{\rm signal} /dt \right]_{\rm spin}$, the signal induced by the mixing is expected to be much larger than the spin-induced ones.
Note, however, that if each magnon event could be detected in other ways, i.e, without the use of cavity, the hidden photon interactions with spin and SM photon may be separately confirmed, which works as strong evidence of hidden photon DM.
Conversely, if the DM signal is discovered in a cavity without magnetic material and the sizable spin-induced signal is also present, one can rule out the hidden photon DM.

\begin{figure}[t]
  \centering
  \begin{tabular}{cc}
    \includegraphics[width=0.48\hsize]{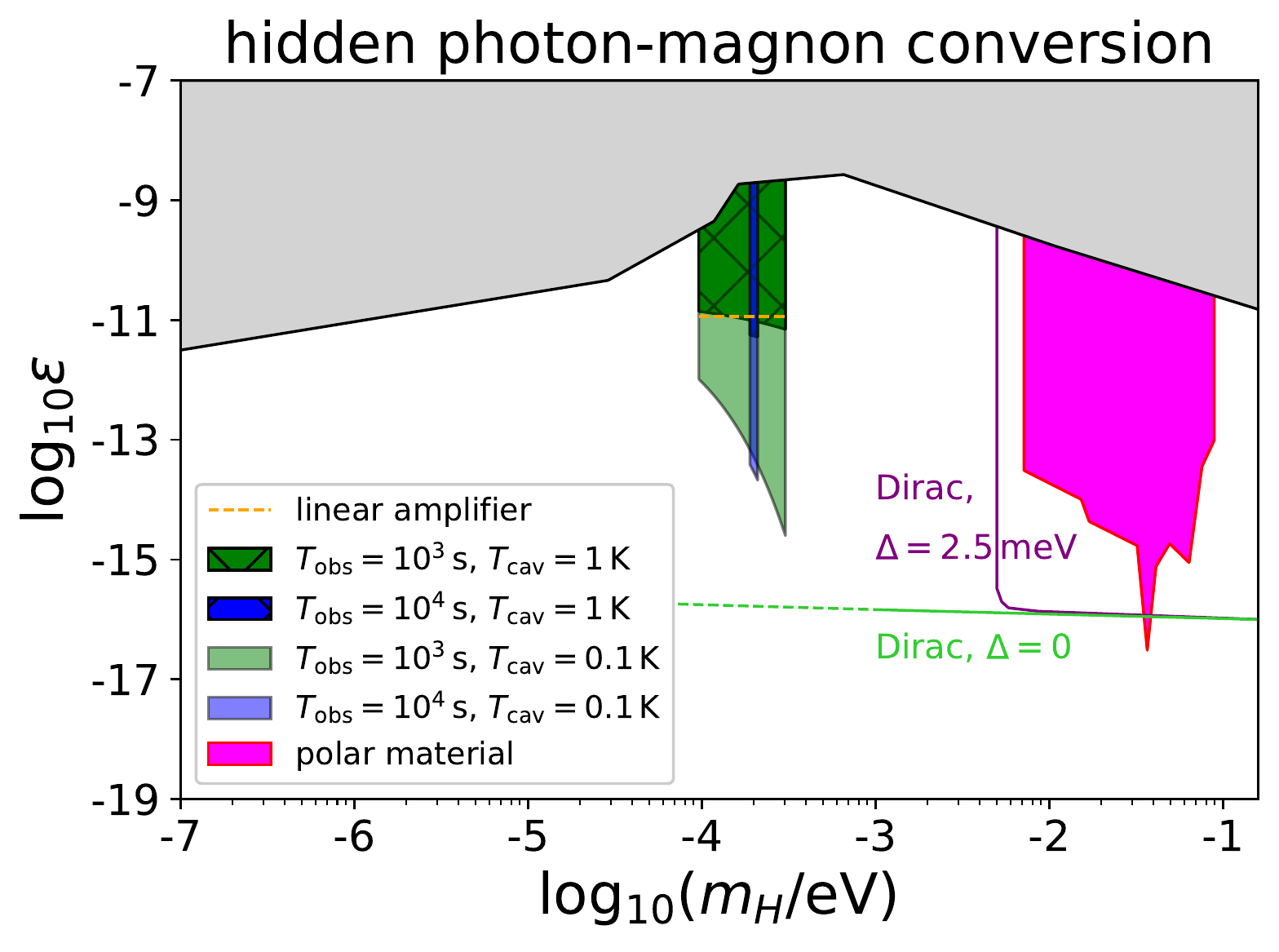}
    \includegraphics[width=0.48\hsize]{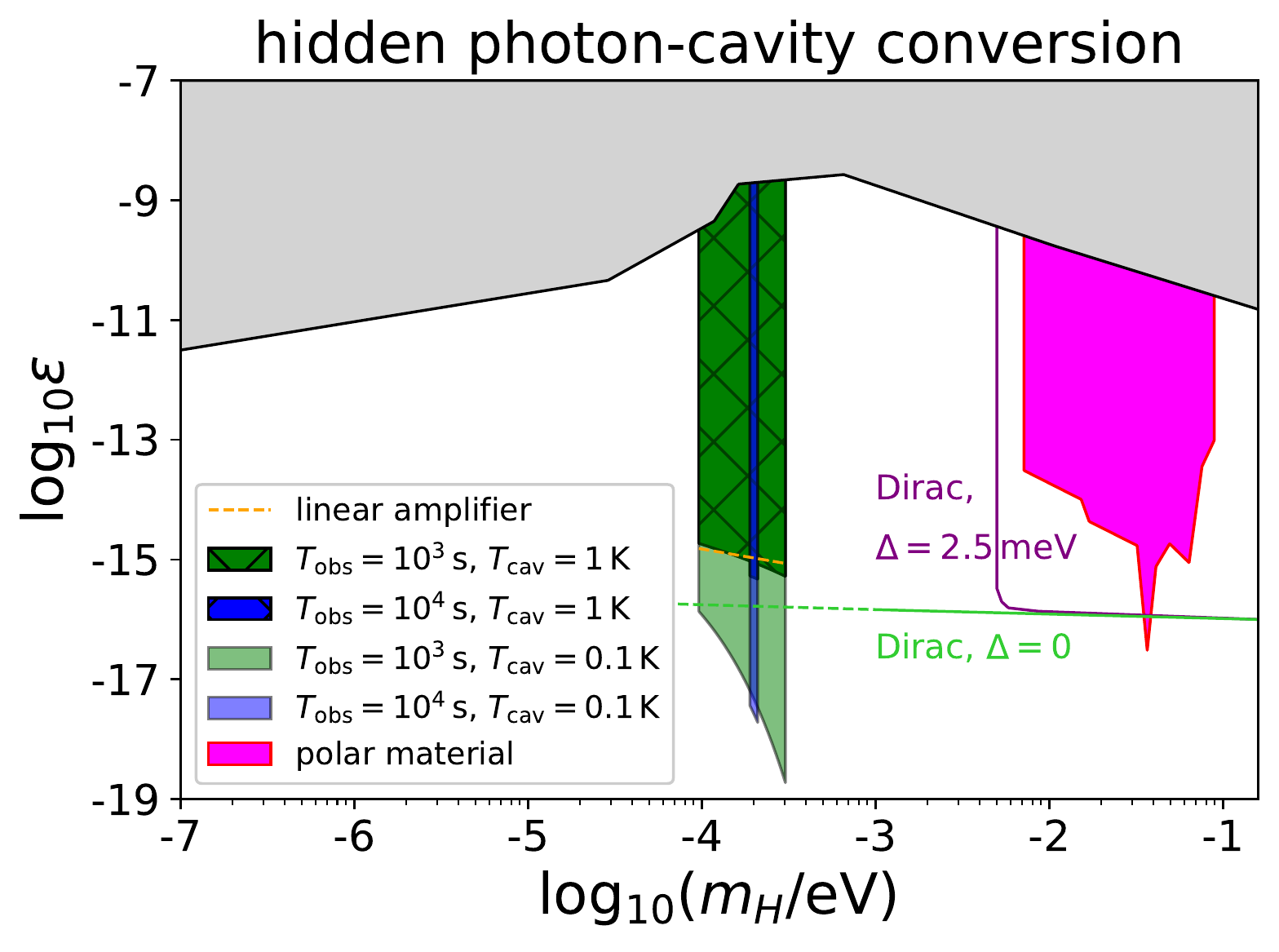}
  \end{tabular}
  \caption{
    Sensitivity of the magnon (left) and cavity (right) detectors in the $m_H$ vs. $\epsilon$ plane.
    We use $M_{\text{target}} = 1\,\text{kg}$ and $T_{\text{total}} = 10\,\text{years}$.
    The other parameters are chosen as $v_H = 10^{-3}$, $\tau = 2\,\mathrm{\mu s}$, and $\sin^2 \theta = \sin^2 \varphi = 1/2$.
    The green and blue colors correspond to an ideal setup case with $(\mathrm{SNR})=3$ and $T_{\text{obs}} = 10^3\,\mathrm{s}$ and $10^4\,\text{s}$, respectively, and the dark-meshed and the light regions show those with $T_{\mathrm{cav}} = 1\,\mathrm{K}$ and $0.1\,\mathrm{K}$, respectively.
    The orange dashed lines correspond to a realistic setup with $T_{\mathrm{obs}} = 10^3\,\mathrm{s}$ and $T_{\mathrm{cav}} \ll m_a$.
    The gray region corresponds to the parameter space already excluded by other experiments.
    Magenta region shows the expected sensitivity of polar materials, while purple and light green lines show that of Dirac materials.}
  \label{fig:sensitivity}
\end{figure}

Let us estimate the experimental sensitivity as done in
Sec.~\ref{sec:sens_axion}.  In Fig.~\ref{fig:sensitivity}, we show the
sensitivity of the magnon (left) and the cavity (right) detectors on the hidden photon with $M_{\text{target}} = 1\,\text{kg}$ and
$T_{\text{total}} = 10\,\text{years}$.
The center of the scan is fixed to be $m_H = 200\,\mathrm{\mu eV}$.
To derive the sensitivity, we use the
parameter choices $v_H = 10^{-3}$, $\tau = 2\,\mathrm{\mu s}$, and
$\sin^2 \theta = \sin^2 \varphi = 1/2$.
For an ideal setup, we again use two different choices of the observation time
$T_{\text{obs}} = 10^3\,\mathrm{s}$ (green) and $10^4\,\mathrm{s}$
(blue), while the dark-meshed and light regions show the sensitivities with $T_{\mathrm{cav}} = 1\,\mathrm{K}$ and $T_{\mathrm{cav}} = 0.1\,\mathrm{K}$, respectively.
The orange dashed lines show the sensitivities of a realistic setup with $T_{\mathrm{obs}} = 10^3\,\mathrm{s}$ and $T_{\mathrm{cav}}\ll m_a$.
Also shown in gray color is the parameter
region already excluded~\cite{McDermott:2019lch}; this includes
constraints from spectral distortions \cite{Arias:2012az},
modifications to $N_{\text{eff}}$~\cite{Arias:2012az}, and stellar
cooling~\cite{An:2013yfc, Redondo:2013lna, Vinyoles:2015aba}.  The magenta
region shows the expected sensitivity using polar materials with
phonon excitation by the hidden photon
absorption~\cite{Knapen:2017ekk}.
The purple (light green) solid line shows the expected sensitivity using Dirac materials with a band gap of $\Delta = 2.5\,\mathrm{meV}$ ($\Delta = 0$) \cite{Hochberg:2017wce}, while the light green dotted line is an extrapolation of the sensitivity assuming that the electron excitation with energy of $\mathcal{O} (10^{-4})\,\mathrm{eV}$ can be detected.
From the figure, we can see the
strong potential of this setup on the hidden photon search.  Even if
we use a much shorter value of $T_{\text{obs}}$ than the canonical
value adopted in the QUAX proposal, a much stronger bound on the
kinetic mixing $\epsilon$ is obtained than the existing ones.  For
models with kinetic mixing between the photon and hidden photon DM, we
can see that the cavity mode can cover a larger parameter region.  It
is notable that the magnon mode can also reach a parameter region
which has not been explored yet.
If one can separate the cavity signal and magnon-induced signal, it is in principle possible that
the hidden photon DM scenario is confirmed by looking at the ratio of both the signals,
although it might be challenging due to the weakness of the magnon signal.

Although we have fixed the central value of $m_H$ for the scan to be $200\,\mathrm{\mu eV}$, the choice of this value is not mandatory.
The mass range to which this search method can be applied is estimated as follows.
As for the heavier region, the strength of the magnetic field will put an upper bound on the applicability.
The hidden photon mass of $200\,\mathrm{\mu eV}$ already corresponds to the magnetic field of $1.7\,\text{T}$, which should be amplified linearly as considering heavier mass.
Thus a few times $10^{-4}\,\mathrm{eV}$ is considered to be the largest mass that can practically be probed.
On the other hand, for the lighter region, the energy deposited to the detector from the hidden photon becomes comparable to the thermal noise when $m_H \sim 10^{-5}\,\mathrm{eV}$.
This lower bound, however, may be loosened by further cooling the detector, though larger cavity may be needed to detect magnon-polariton as a final state particle.

So far, we have seen that the cavity mode has a better sensitivity
than the magnon mode if the coupling to the hidden photon is dominated
by the kinetic mixing given in Eq.\ \eqref{L_hiddenphoton}.  This
conclusion is, however, model dependent.
For example, one can consider a model where the electron has only the magnetic interaction with the hidden photon:
\begin{align}
  \mathcal L = -\frac{1}{4} H_{\mu\nu}H^{\mu\nu} - \frac{1}{4} B_{\mu\nu}B^{\mu\nu}
  +\frac{1}{2} m_H^2 H_\mu H^\mu + \frac{m_e}{M^2} \overline\psi\sigma_{\mu\nu} \psi H^{\mu\nu},
\end{align}
with some cutoff scale $M$. In this case, the magnon-induced photon is practically the only possible signal and the strongest constraint on $\epsilon$, which is now reinterpreted as the effective interaction strength $m_e/M^2$, may be obtained from the magnon excitation.

\section{Conclusions and discussion}
\label{sec:conc}

We have shown that the light boson DM (axion and hidden photon) can be converted into magnon and it can be used as a DM detection method. Such an idea was already given in the QUAX proposal~\cite{Barbieri:2016vwg} for the axion DM detection and we have shown that a similar process happens for the hidden photon DM.
A key observation is that the hidden photon has a magnetic interaction with electrons, which induces a spin wave or the magnon in the ferromagnetic/ferrimagnetic insulator. Since the magnon dispersion relation can be adjusted by applying the external magnetic field, one can scan the hidden photon DM mass.
Unfortunately, such a spin-induced signal is smaller than the conventional hidden photon to SM conversion in the cavity, but it can be used to distinguish the axion DM (DFSZ or flaxion model) and hidden photon DM since the former predicts relatively large signal from the DM-magnon interaction.

Below we comment on ideas of hidden photon DM detection in the condensed-matter system.
Refs.~\cite{Hochberg:2016ajh,Hochberg:2016sqx} considered superconductor and semiconductor as a target material. The hidden photon is absorbed by electrons in the conducting band and it emits the (acoustic) phonon, hence it is a scattering process $\vec H + e \to P + e$, where $P$ denotes the phonon.
Refs.~\cite{Knapen:2017ekk,Griffin:2018bjn} also considered the hidden photon absorption by polar material, which has gapped optical phonon modes. It may be regarded as a hidden photon conversion process into an optical phonon, followed by the dissipation of the optical phonon.
On the other hand, we focused on the hidden photon conversion into magnon in a ferromagnetic or ferrimagnetic insulator.

We have only considered resonant conversion into the magnon. It is rather regarded as a magnon-polariton so that the magnon effectively induces a cavity photon mode~\cite{Barbieri:2016vwg}. While the conversion rate is enhanced, one drawback of this idea is that it takes a long time to scan the wide range of DM mass.
It is in principle possible that the magnon decays into several quanta, such as two magnons~\cite{Kreisel:2009,Ruckriegel:2014} or magnon plus phonon~\cite{Streib:2019}. Multi-phonon processes in the context of light DM detection was discussed in Refs.~\cite{Schutz:2016tid,Knapen:2016cue,Acanfora:2019con} for superfluid helium target and in Ref.~\cite{Campbell-Deem:2019hdx} for crystal target. In such a case the kinematical constraint is weakened and wide mass range may be covered while the excitation rate is suppressed.
We keep a detailed study of this issue as a future work~\cite{Chigusa:2020}.

Here we point out that other ideas for axion DM detection may also be used as a hidden photon detector.
In Ref.~\cite{Marsh:2018dlj} a novel method to detect axion DM was proposed using the topological antiferromagnet insulator.
The axion is assumed to have an interaction with photon through the Chern-Simons term like (\ref{LTI_axion}).
In a topological magnetic insulator, the magnon may also have a similar Chern-Simons coupling to the photon~\cite{Li:2010}.
Under the applied magnetic field, the background DM axion is converted into the electric field. It is again converted into the magnon under the magnetic field, which induces photon emission due to the boundary effect.
By choosing the magnetic field appropriately, the intermediate magnon hits the resonance to enhance the signal.
Notably, the same idea also applies to the hidden photon DM. The hidden photon DM is converted to the electric field through the kinetic mixing
\begin{align}
	\mathcal L = \frac{\epsilon}{2} H_{\mu\nu} F^{\mu\nu} = -\epsilon(\vec E_H\cdot \vec E- \vec B_H\cdot \vec B).
	\label{LTI_hidden}
\end{align}
Given $|\vec E_H| \gg |\vec B_H|$, the hidden photon DM mainly produces electric fields. Then it is converted into the magnon under the magnetic field, as explained above.\footnote{
	In the mass eigenstate basis $H_\mu''$, one can interpret the same process as a result of direct interaction of the magnon with the photon and hidden photon through the Chern-Simons term.
}
Comparing (\ref{LTI_axion}) with (\ref{LTI_hidden}), one obtains a sensitivity on the kinetic mixing parameter $\epsilon$ by replacing the sensitivity on $(m_a,f)$ obtained in Ref.~\cite{Marsh:2018dlj} through the correspondence
\begin{align}
	\epsilon = C_{a\gamma}\frac{\alpha_e}{2\pi} \frac{B_0}{m_a f}
	\simeq 2\times 10^{-17}\,C_{a\gamma}\left( \frac{10^{10}\,{\rm GeV}}{f} \right)\left( \frac{1\,{\rm meV}}{m_a} \right)
	\left( \frac{B_0}{1\,{\rm T}} \right),
\end{align}
where $B_0$ is the applied magnetic field. Thus it may have a very good sensitivity on the hidden photon DM.
We will also come back to this issue in a separate publication.

\section*{Acknowledgments}

This work was supported by JSPS KAKENHI Grant (Nos. 17J00813 [SC], 16H06490 [TM],
18K03608 [TM], 18K03609 [KN], 15H05888 [KN] and 17H06359 [KN]).

\appendix

\section{Effective Hamiltonian of magnon}
\label{sec:heff}

Here, we derive the magnon couplings to axion and hidden photon,
starting from the Lorentz-invariant quantum field theory (QFT).  Let
us denote the electron field operator in the QFT as
\begin{align}
  \psi (x) = \int \frac{d^3 p}{(2\pi)^3 \sqrt{2p_0}}
  \sum_s a_{\vec{p},s} u_{\vec{p},s} e^{-ipx},
\end{align}
where $p_0\equiv\sqrt{\vec{p}^2+m_e^2}$ and
\begin{align}
  u_{\vec{p},s} =
  \left( \begin{array}{c}
      \sqrt{p_0+m} \, \chi_s \\
      \sqrt{p_0-m} \, \vec{e}_p \cdot \vec{\sigma} \chi_s
    \end{array}
  \right),
\end{align}
with $\vec{e}_p$ denoting the unit vector pointing to the direction of
$\vec{p}$, and $\chi_s=(1,0)^T$ or $(0,1)^T$.  (We adopt the Dirac
representation of the $\gamma$-matrices.)  Besides, the creation
and annihilation operators of the electron are denoted as
$a_{\vec{p},s}$ and $a^\dagger_{\vec{p},s}$, respectively, which
satisfy $\{a_{\vec{p},s}, a^\dagger_{\vec{p}',s'}\} =(2\pi)^3\delta
(\vec{p}-\vec{p}')\delta_{ss'}$.  We are interested in the system
containing only the electron, so we neglect the positron degrees
freedom.  Concentrating on non-relativistic degrees of freedom, the
spin operator in the QFT is given by
\begin{align}
  \vec{S}^{\rm (QFT)} =
  \int \frac{d^3 p}{(2\pi)^3} \sum_{s, s'}
  a^\dagger_{\vec{p},s} a_{\vec{p},s'}
  \chi_s^\dagger \vec{\sigma} \chi_{s'}.
\end{align}
Notice that the spin operator given above satisfies the relevant
commutation relations.

We expect that the total spin operator in the QFT is matched to that
in the Heisenberg model as
\begin{align}
  \vec{S}^{\rm (QFT)} \rightarrow \sum_\ell \vec{S}_\ell.
\end{align}
Hereafter, we derive the effective interaction of the magnon using
this matching condition as well as assuming the locality of the
interaction.

In the model with the hidden photon, the magnon couples to the hidden
photon via the kinetic mixing with the hypercharge photon given in
Eq.\ \eqref{L_hiddenphoton}.  Because we are interested in the energy
scale much lower than the electron mass, we can concentrate on the
effective field theory that contains only non-relativistic electron,
photon, and hidden photon.  In such a case, the only relevant
interaction of the hidden photon with the electron is from the mixing
of the hidden photon with the ordinary photon, and is given by
\begin{align}
  {\cal L} = -\epsilon e Q H_\mu \bar{\psi} \gamma^\mu \psi,
\end{align}
where $Q$ is the charge of the electron (in units of $e$), and
$\epsilon \equiv \epsilon_Y c_W$.   Then, we can find
\begin{align}
  H_{\rm int} \simeq
  \frac{\epsilon eQ}{2m_e} \epsilon_{ijk} \int d^3x
  \frac{d^3 p}{(2\pi)^3} \frac{d^3 p'}{(2\pi)^3} \sum_{s,s'}
  a^\dagger_{\vec{p},s} a_{\vec{p}',s'}
  (\partial_i H_j) e^{i(p-p')x} \chi^\dagger_s \sigma^k \chi_{s'}
  +
  \epsilon \int d^3 x H_\mu j^\mu ,
  \label{H_int(hiddenph)}
\end{align}
where
\begin{align}
  j^0 \equiv &\,
  - e Q
  \int \frac{d^3 p}{(2\pi)^3} \frac{d^3 p'}{(2\pi)^3} \sum_{s}
  a^\dagger_{\vec{p},s} a_{\vec{p}',s}
  e^{i(p-p')x},
  \\
  \vec{j} \equiv &\,
  - e Q
  \int \frac{d^3 p}{(2\pi)^3} \frac{d^3 p'}{(2\pi)^3} \sum_{s}
  \frac{\vec{p}+\vec{p'}}{2m_e}
  a^\dagger_{\vec{p},s} a_{\vec{p}',s}
  e^{i(p-p')x}.
\end{align}
The first term of the right-hand side gives the coupling of magnon to
the hidden photon.  Assuming the locality of the interaction, and
using the (discrete) translational invariance of the system, we expect
that the effective interaction between the spin (i.e., magnon) and the
hidden photon contains the following term:
\begin{align}
  H_{\rm int} \ni - \frac{\epsilon eQ}{m_e}
  \sum_\ell \vec{B}_H (\vec{x}_\ell)\cdot \vec{S}_\ell,
\end{align}
with $\vec{B}_H$ being the hidden magnetic field.  The second term of
the right-hand side of Eq.\ \eqref{H_int(hiddenph)} is the coupling of
the (ordinary) current with the vector potential of the hidden photon.
So far, we have concentrated on the coupling between the hidden photon
and the electron.  The hidden photon coupling to the nucleon can be
also derived similarly. The nucleon counterpart of the second term of Eq.\ \eqref{H_int(hiddenph)} may cause the hidden
photon-phonon conversion. However, it is not kinematically allowed
unless the (optical) phonon energy gap happens to be close to the
hidden photon mass. The hidden photon absorption by polar material was
considered in Refs.~\cite{Knapen:2017ekk,Griffin:2018bjn} for the mass
range of $10^{-2}$--$10^{-1}$\,eV. We consider lighter DM mass region
around meV, so we neglect such an effect.\footnote{ Absorption of
  hidden photon DM as light as meV by the Dirac material was
  considered in Ref.~\cite{Hochberg:2017wce}.  }

The magnon-axion coupling originates from the axion-electron
interaction (see the last term of Eq.\ \eqref{L_DFSZ}).  Using the
fact that, in the non-relativistic limit,
\begin{align}
  \vec{S}^{\rm (QFT)} \simeq
  \frac{1}{2}
  \int d^3 x \bar{\psi} \vec{\gamma} \gamma_5 \psi,
\end{align}
we obtain the coupling of an axion to a magnon as
\begin{align}
  H_{\rm int} = \frac{1}{f} \sum_\ell \vec{\nabla} a (x_\ell) \cdot \vec{S}_\ell.
\end{align}

\section{Classical calculation of conversion rate}  \label{sec:classical}

Let us reproduce the same result with classical calculation~\cite{Barbieri:2016vwg}.
We treat the magnetization $\vec M$ of the material as a classical magnetic moment and study its motion under the classical axion background.
Neglecting damping effects due to radiation, spin-spin or spin-lattice interactions, the classical equation of motion is given by
\begin{align}
	\dot{\vec M} =\frac{e}{m_e} \vec M \times \vec B,~~~~~~\vec B = B_z^0 \vec e_z + \vec B_a.
\end{align}
We find
\begin{align}
	&\ddot M_x + \omega_L^2 M_x = \frac{e M_z}{m_e} \left( \omega_L B_x^a- \dot B_a^y \right),\\
	&\ddot M_y + \omega_L^2 M_y = \frac{e M_z}{m_e} \left( \omega_L B_y^a+ \dot B_a^x \right).
\end{align}
We can rewrite these equations by using $M_{\pm}\equiv M_x\pm i M_y$ as
\begin{align}
	&\ddot M_+ + \omega_L^2 M_+= i \frac{esN m_a^2 a_0 v^+}{m_e}e^{-i(m_at + \delta)}, \\
	&\ddot M_- + \omega_L^2 M_-= -i \frac{esN m_a^2 a_0 v^-}{m_e}e^{i(m_at + \delta)}.
\end{align}
The solution is
\begin{align}
	M_+(t) = \frac{i\widetilde V e^{-i(m_at+\delta)}}{m_a^2-\omega_L^2}\left[
		-1 + e^{i m_at}\left(\cos(\omega_L t)-i \frac{m_a}{\omega_L}\sin(\omega_L t)\right)
	\right] + \frac{i\widetilde V \sin(\omega_Lt)\sin\delta}{m_a \omega_L},
\end{align}
where
\begin{align}
	\widetilde V \equiv \frac{esN m_a^2 a_0 v_a^+}{m_e f}.
\end{align}
Taking the limit $\omega_L = m_a$, we obtain
\begin{align}
	M_+(t) \simeq \frac{i\widetilde V t e^{-i(m_at+\delta)}}{2m_a},~~~~~~M_-(t) \simeq -\frac{i\widetilde V^* t e^{i(m_at+\delta)}}{2m_a},
\end{align}
for $m_a t \gg 1$.
The power obtained by the axion wind is then estimated as
\begin{align}
	\frac{dE_{M}}{dt} = \frac{d}{dt}(\vec B\cdot \vec M)
	\simeq \frac{sN \rho_{\rm DM} m_a (v_a^{x2}+v_a^{y2})t}{f^2} \sin^2(m_at+\delta).
\end{align}
After averaging $\sin^2(m_at+\delta)=1/2$, it is consistent with (\ref{power_magnon}).



\end{document}